\documentclass[twoside,11pt]{article}

\usepackage{jmlr2e}
\usepackage{multirow}
\jmlrheading{}{2016}{}{}{}{Hadi Fanaee-T and Joao Gama}

\ShortHeadings{SimTensor: A synthetic tensor data generator}{Fanaee-T and Gama}
\firstpageno{1}

\begin{document}

\title{SimTensor: A synthetic tensor data generator}

\author{\name Hadi Fanaee-T \email hadi.f.tork@inesctec.pt \\
       \addr INESC TEC\\
       Campus of FEUP\\
       4200-465 Porto, Portugal
       \AND
       \name Joao Gama \email jgama@fep.up.pt \\
       \addr INESC TEC\\
       Campus of FEUP\\
       4200-465 Porto, Portugal}

\editor{}

\maketitle

\begin{abstract}%   <- trailing '%' for backward compatibility of .sty file
SimTensor is a multi-platform, open-source software for generating artificial tensor data (either with CP/PARAFAC or Tucker structure) for reproducible research on tensor factorization algorithms. SimTensor is a stand-alone application based on MATALB. It provides a wide range of facilities for generating tensor data with various configurations. It comes with a user-friendly graphical user interface, which enables the user to generate tensors with complicated settings in an easy way. It also has this facility to export generated data to universal formats such as CSV and HDF5, which can be imported via a wide range of programming languages (C, C++, Java, R, Fortran, MATLAB, Perl, Python, and many more). The most innovative part of SimTensor is this that can generate temporal tensors with periodic waves, seasonal effects and streaming structure. it can apply constraints such as non-negativity and different kinds of sparsity to the data. SimTensor also provides this facility to simulate different kinds of change-points and inject various types of anomalies. The source code and binary versions of SimTensor is available for download in http://www.simtensor.org. 
\end{abstract}


\section{Introduction}

Tensor factorizations have lots of applications in data mining, machine learning, and signal processing as well as chemometrics, and computer vision \citep{kolda2009tensor,morup2011applications,fanaee2016tensor,papalexakis2016tensors}. Obviously any improvement in the accuracy of tensor factorization algorithms advances the state-of-the-art in the above-mentioned domains. Developing accurate algorithms for tensor factorization require benchmark data sets with various realistic adjustable configurations. One of the characteristics that is not included in many existing tensor data generators is time-changing behavior of realistic tensors as well as many real issues such as change-points and anomalies. The objective of SimTensor is to provide a standard and comprehensive framework for generating synthetic tensor structured data with focus on the time-changing characteristics of data. The SimTensor is developed based upon of many available codes and techniques in the literature of tensor analysis. So it is benefited a lot from the open-source contributions. Our methodology was to carefully study the available features and considered issues in the existing data generators and the presented approaches in the literature for creating test problems. Besides, some ideas such as injecting anomalous tensor is being presented for the first time in SimTensor and is not reported elsewhere. Therefore, SimTensor can be considered as the state-of-the-art tool for generating synthetic tensor data with either CP \citep{carroll1970analysis,carroll1980candelinc} or Tucker \citep{tucker1966some} structure with focus on time-changing tensors that are very realistic in many applications. 

The existing data generators featured in toolboxes such as Tensor Toolbox \citep{TTB_Software} and N-Way Toolbox \citep{andersson2000n} are presented as functions in MATLAB. Hence the access of users of other programming environments is limited since MATLAB is a licensed software. The SimTensor provides a stand-alone solution for wider range of practitioners who develop tensor factorization algorithms in various programming languages and operating systems and want to test their algorithms with several realistic configurations. Therefore, it is expected that SimTensor can contribute to reproducible research on tensor factorization algorithms. It is also open-source, so the community can actively contribute to the project and add their desired features to it according to the requirements that are not considered in the current version.

\section{Generating Non-temporal Random Factors}
One of the basic step to generate synthetic tensors with CP structure is to create random factor matrices $U^{(n)}$ and then compute sum of multiway-way outer products:

\begin{equation} \label{CP}
 x_{ijk} =  \sum_{r=1}^{R}u_{ir}^{(1)}u_{jr}^{(2)}u_{kr}^{(3)}. \end{equation}
 
 Where \textit{R} is the number of components, $I, J, K$ represent the size of tensor in each dimension, and N is order of tensor. If we assume that columns of $U^{_(n)}$ are normalized we can introduce the vector $\lambda \in \mathbb{R}^{R} $  and re-write (\ref{CP}) as:
 
 \begin{equation}   
   x_{ijk} =  \sum_{r=1}^{R} \lambda_r u_{ir}^{(1)}u_{jr}^{(2)}u_{kr}^{(3)}.
   \end{equation}

For the Tucker tensor in addition to factor matrices we have to generate a random core tensor $\mathcal{G}$ as well. For instance, for a third-order tensor we have:

 \begin{equation}   
   x_{ijk} =  \sum_{r_1=1}^{R_1}  \sum_{r_2=1}^{R_2}  \sum_{r_3=1}^{R_3} g_{r_1r_2r_3} u_{ir_1}^{(1)}u_{jr_2}^{(2)}u_{kr_3}^{(3)}.
   \end{equation}
  
In order to generate $U^{(n)}$ we can use different approaches. These techniques will be described in the following subsections.

\subsection{Gamma Distribution}
In SimTensor, the module $gamma$ generates random factors from  $\Gamma (k, \theta)$ where $\Gamma$ is Gamma function, $k$ is  positive scalar value called shape parameter, and $\theta$ is nonnegative scalar value called scale parameter. As \citep{hu2015scalable} in SimTensor we choose $k$ from $|{\mathcal {N}}(\mu ,\,\sigma ^{2})|$ where $\mathcal {N}$ is Gaussian distribution, $\mu$ is mean, and $\sigma ^{2}$ is variance. In SimTensor, by default $\mu$ and $\sigma ^{2}$ are set as 0.1 and $\theta$ is set as 0.01.

\subsection{Multivariate Gaussian Distribution}
The module $multi\_normal\_dist$ allows to generate $R$ columns in factor matrix via multivariate Gaussian distribution with different $\mu$ and $\sigma$ parameters \citep{zhao2016bayesian} for each column. In SimTensor by default $\mu$ and $\sigma$ are set as 0 and 1, for all columns. 

\subsection{Uniformly Distributed Random Numbers}
The module $rand$ in SimTensor generates random factors with uniformly distributed random numbers on [0,1]. This method is widely used for generating random factors, for instance in Tensor toolbox \citep{TTB_Software} and elsewhere.

\subsection{Standard Normal Distribution}
The module $randn$ generate factor matrices drawn from the standard normal distribution with $\mu=0$ and $\sigma=1$ for all columns. Note that in the current version of SimTensor $randn$ is identical to module $multi\_normal\_dist$. Because by default we set $\mu=0$ and $\sigma=1$ in all columns for both methods. However, $multi\_normal\_dist$ opposed to $randn$ has this ability to generate columns in factor matrices with different $\mu$ and $\sigma$. The $randn$ module is also used in Tensor toolbox \citep{TTB_Software} for creating test problems. 

\subsection{Random Orthogonal Matrices}
The module $orthogonal$ is borrowed from Tensor toolbox \citep{TTB_Software} and generates a random $n \times n$ orthogonal matrix, a matrix distribution uniform over the manifold of orthogonal matrices with respect to the induced $\mathbb{R}^{n^2}$ Lebesgue measure \citep{RandOrthMat}. 

\subsection{Stochastic}
The module $stochastic$ as its equivalent in Tensor toolbox \citep{TTB_Software} generates nonnegative factor matrices so that each column sums to one. This method works as follow. First, a factor matrix of $U^{n} \in \mathbb{R}^{I^{n} \times R^{n}}$ is generated via uniformly distributed random numbers on [0,1]. Then the sum of each column of $U^{n}$ is obtained and stored in vector $S$. Final generated matrix is obtained by multiplying $U^{n}$ to diagonal matrix of $S^{-1}$.

\subsection{Binary factors}
The idea of binary factors is borrowed from work of \citep{Trimine} that creates boolean synthetic tensor by generating random binary factors. The method is very simple. First, we generate matrix of $U^{n} \in \mathbb{R}^{I^{n} \times R^{n}}$ filled with zero. Then for each row of matrix we randomly fill one column of matrix with one, such that in each row we will have only one column with one.

\subsection{Random Core Tensor/Lambda Generator}
SimTensor allows the user to generate random core/tensor (for Tucker) and lambda (for CP). The methods include vector of ones (ones), uniformly distributed random numbers (rand), and standard normal distribution (randn). The user also can customize the generated vector. For Tucker case the generated vector is automatically transformed to the core tensor. 

\section{Generating Temporal Random Factors}
SimTensor enables the user to create temporal factors with different strategies such as periodic waves, seasonal effects, and streaming. 

\subsection{Periodic waves}
In SimTensor is possible to simulate periodic waves such as Cosine, Sine, as well as Square and Sawtooth waves. The user is able to determine the number of waves and frequency of waves in each setting. The idea is inspired by \citep{lemm2011introduction} for generating artificial data for testing algorithms for factroization of shift-invariant multilinear data which is a real case in neuroimaging data. 

\subsection{Seasonal effects}
Seasonal effects are very realistic for those data sets that are generated by humans. In such data sets there is a repeating pattern during specific time interval such as days, weeks, seasons, and so forth. For instance, in disease surveillance data \citep{fanaee2015eigenevent} disease outbreaks such as Influenza are more frequent in winters. In computer networks different traffic usage pattern can be detected. For instance, it is observed that traffic peak occurs around mid-day and during the night \citep{tune2013internet}. In transportation systems the peak of traffic is on the morning when people go to work and evening when they back home \citep{tan2013new}. There are lots of examples of such temporal seasonality. In SimTensor is possible to simulate different seasonal effects with various temporal granularities.  Sometimes also in some other applications such as animal migration we may find some specific seasonality patterns which can be different from human-centered systems. In SimTensor the user is able to define some customized seasonal effects like this as well. The user also is able to define a particular growth rate. In this case, the data items in each cycle will be multiplied by a weight. 

As in many realistic scenarios (e.g. in transportation systems) also we may experience multiple seasonality patterns in data. It is extremely easy to simulate such scenarios in SimTensor with defining multiple factors with different seasonal grounds.   

\subsection{Streaming tensor}
In the approach that is exploited in \citep{nion2009adaptive} for generating time-varying factor matrices it is assumed that data arrives in a streaming fashion and it is expected that data in \textit{t+1} is a sample of data of \textit{t} with some little changes. This change is controlled by a parameter that is called \textit{variation control parameter} and can be defined by the user. This can be used for evaluation of any incremental or streaming tensor analysis algorithm  \citep{sun2008incremental,zhouaccelerating,fanaee2015multi}.

\section{Simulating Change Points}
Shifts and drifts are realistic characteristics in many applications. It is important to assess the performance of tensor factorization algorithms in dealing with such changes. In SimTensor it is possible to create artificial change points in the factor matrices related to the temporal dimension. Depending on the defined period by the user different types of changes can be simulated, including temporary changes, structural shifts, and singular outliers. For instance, if temporal mode has size of 100, the user can simulate a shift by defining the change point with start point of 51 and end point of 100. Example of  short-term changes (or events) can be start point of 20 and end point of 25. And finally if user is interested to simulate singular changes (or temporal outliers) she can define start/end point at the same time instant. 

\section{Simulating Anomalies}
Those patterns in data that do not conform to expected behavior are called anomalies \citep{chandola2009anomaly,fanaee2016tensor}. In different contexts anomalies may have diverse interpretations. In medical domain, for instance it can be translated to disease outbreak. In industrial settings it might be a fault. Or in transportation systems it might be translated as a events \citep{fanaee2016event}. However, anomalies are different from outliers, in the sense that anomalies occur in a systematic way. In fact anomalies are generated by anomalous process while outliers can be only small errors in information systems or data gathering \citep{fanaee2014event}. 
In SimTensor we for the first time propose a new method for simulating anomalies. The idea is that we first create a smaller random tensor with CP structure and then inject it inside the bigger generated tensor. Actually we replace a small portion of the original tensor with the new generated tensor. This will be a challenge for tensor factorization algorithms to discover this small injected tensor. It is assumed that those approaches that can detect the smaller tensor should perform well for detecting anomalies from tensor-structured data. 

\section{Simulating Noise}

Noise is the inherent part of many data sets. Three methods are available in SimTensor for simulating noises. The user is able to apply noise directly to the factor matrices; apply an additive white Gaussian noise with controllable noise level on the final generated tensor (more common) via the method of \citep{Viswanathan}; or apply sparse white noise on the final tensor.

\section{Constraints on factor matrices}
SimTensor enables the user to add non-negativity constraint in two ways, either on factor matrices or on the final tensor. The user can also apply different constraints such as correlation between columns in the factor matrices on the factor matrices or angle between columns of the factor matrices. Note that this can be applied to non-temporal modes. It is also possible to normalize the final tensor or perform a sign fix operation on the final tensor.

\section{Generating Sparse tensors}

Three mechanisms are included for generating sparse tensors. The first one is to apply sparsity constraint on the factor matrices by randomly removing non-zero elements from the dense created factor matrices. The second strategy is to generate CP or Tucker tensor with the dense factor matrices and then remove some random non-zero elements. The third methodology is to create sparse count tensors based on the idea of \citep{hu2015scalable}. In this method we firstly generate random factors with Gamma random numbers. Then we create CP tensor with the generated random factors and finally feed this tensor as input parameters for generating tensor with Poisson distribution. This kind of synthetic data can be used for evaluation of Bayesian tensor factorization algorithms \citep{hu2015scalable}.

\begin{table}
\scalebox{0.7}{
    \begin{tabular}{|p{2.5cm}|p{3.5cm}|p{7cm}|p{7cm}|}
    \hline
    Component                     & Module                                   & Target tensor factorization algorithm                               & Target Data                                                             \\ \hline
    \multirow{4}{3cm}{Non-temporal factor generator} & rand, randn,  multi\_normal\_dist & Traditional approaches, for example CP-ALS \citep{carroll1970analysis,carroll1980candelinc} or Tucker-ALS \citep{tucker1966some} & High quality multi-linear data                                                                          \\ \cline{2-4}
                                & Orthogonal                                    & HOSVD \citep{de2000multilinear}, HOOI \citep{de2000best} or similar                   &  High quality multi-linear data                                                                         \\ \cline{2-4}
                                  & Stochastic                             & Non-negative  algorithms like \citep{carroll1989fitting,kim2007nonnegative}       & Visual data (e.g. image or video)  and count data (e.g. recommender systems)      \\ \cline{2-4}
                                  & Binary                                   & Boolean algorithms such as \citep{miettinen2011boolean}                 & Social network data                                                             \\ \hline
    \multirow{4}{3cm}{Temporal factor generator}     & Periodic waves                           & Shift-invariant TD algorithms like ShiftCP \citep{morup2008shift}            & Neuroimaging data (e.g. EEG)                                                    \\  \cline{2-4}
                                 & Seasonal effects                         & All Algorithms                                          & Human generated data sets (e.g. Internet traffic, mobility data, etc.) \\  \cline{2-4}
                                 & Streaming                                & Online Algorithms like PARAFAC-SDT \citep{nion2009adaptive}, OnlineCP \citep{zhouaccelerating} or similar   & Streaming data                                                                  \\  \cline{2-4}
                                 & Change-points                            & Incremental algorithms such as  DTA, STA, or WTA \citep{sun2008incremental}           & Temporal tensors                                                                   \\ \hline
    \multirow{4}{3cm}{Effects}                          & Anomalies                                & All algorithms (with focus on the evaluation of model's discriminability)                                         & Data sets with structural anomalies                                           \\ \cline{2-4}
                                 & Noise                                    & All algorithms (with focus on the evaluation of model's power)                                         & All Data sets                                                     \\ \cline{2-4}
                                 & Non-negativity constraints                 & Non-negative  algorithms like \citep{carroll1989fitting,kim2007nonnegative}                      & Visual data (e.g. image or video)  and count data (e.g. recommender systems)      \\ \cline{2-4}
                                 & Sparsity constraints                      & Sparse-friendly algorithms such as \citep{kolda2008scalable}             & Incomplete data sets                \\ \cline{2-4}
                                 & Sparse Count Tensors                      & Sparse Bayesian  algorithms such as \citep{hu2015scalable}                          & Large-scale sparse count data in recommender systems and social networks               \\ \hline
    \end{tabular}
    }
      \caption{A guide to choose the the component and module in SimTensor for evaluation of a specific tensor factorization algorithm or testing an algorithm in a particular application.}\label{table:apps}
\end{table}

\section{How to use SimTensor?}
Depending on the evaluation objective SimTensor can generate synthetic data for various family of tensor factorization algorithms. It also can create data sets with different realistic characteristics that exists in many applications. In Table \ref{table:apps} a guide is presented to choose the right module and component for generating synthetic tensor when any specific type of algorithms or data sets is desired.

\acks{We would like to acknowledge support for this project
from the European Comission (EU FP7 grant ICT-2013-612944)}. The development of SimTensor has benefited from the tensor analysis community, including the works and the released codes by: Andrzej Cichocki, Anh-Huy Phan, Arun Tejasvi Chaganty, Brett W. Bader, Changwei Hu, Changyou Chen, Christos Faloutsos, Claus A. Andersson, Daniel M. Dunlavy, Dimitri Nion, Eric C. Chi, Evrim Acar, Giorgio Tomasi, Guoxu Zhou, Haesun Park, Ian Davidson, Jackson Mayo, James Bailey, Jimeng Sun, Jingu Kim, Laurent Sorber, Lawrence Carin, Lieven De Lathauwer, Liqing Zhang, Marc Van Barel, Masatoshi Yoshikawa, Matthew Harding, Nguyen Xuan Vinh, Nicholas D. Sidiropoulos, Nico Vervliet, Otto Debals, Papalexakis, Evangelos E., Percy Liang, Petr Tichavsky, Piyush Rai, Prateek Jain, Qibin Zhao, Rafal Zdunekb, Rasmus Bro, Rong Pan, Sammy Hansen, Sewoong Oh, Shun-ichi Amari, Shuo Zhou, Tamara G. Kolda, Tatsuya Yokotaa, Tomoharu Iwata, Volodymyr Kuleshov, Wotao Yin, Xiaomin Fang, Xingyu Wang , Yangyang Xu, Yasuko Matsubara, Yasushi Sakurai, Yu Zhang , Yukihiko Yamashitac, Yunlong He, Yunzhe Jia.


\bibliography{ref}

\begin{thebibliography}{32}
\providecommand{\natexlab}[1]{#1}
\providecommand{\url}[1]{\texttt{#1}}
\expandafter\ifx\csname urlstyle\endcsname\relax
  \providecommand{\doi}[1]{doi: #1}\else
  \providecommand{\doi}{doi: \begingroup \urlstyle{rm}\Url}\fi

\bibitem[Andersson and Bro(2000)]{andersson2000n}
Claus~A Andersson and Rasmus Bro.
\newblock The n-way toolbox for matlab.
\newblock \emph{Chemometrics and intelligent laboratory systems}, 52\penalty0
  (1):\penalty0 1--4, 2000.

\bibitem[Bader et~al.(2015)Bader, Kolda, et~al.]{TTB_Software}
Brett~W. Bader, Tamara~G. Kolda, et~al.
\newblock Matlab tensor toolbox version 2.6.
\newblock Available online, February 2015.
\newblock URL \url{http://www.sandia.gov/~tgkolda/TensorToolbox/}.

\bibitem[Carroll and Chang(1970)]{carroll1970analysis}
J~Douglas Carroll and Jih-Jie Chang.
\newblock Analysis of individual differences in multidimensional scaling via an
  n-way generalization of “eckart-young” decomposition.
\newblock \emph{Psychometrika}, 35\penalty0 (3):\penalty0 283--319, 1970.

\bibitem[Carroll et~al.(1980)Carroll, Pruzansky, and
  Kruskal]{carroll1980candelinc}
J~Douglas Carroll, Sandra Pruzansky, and Joseph~B Kruskal.
\newblock Candelinc: A general approach to multidimensional analysis of
  many-way arrays with linear constraints on parameters.
\newblock \emph{Psychometrika}, 45\penalty0 (1):\penalty0 3--24, 1980.

\bibitem[Carroll et~al.(1989)Carroll, De~Soete, and
  Pruzansky]{carroll1989fitting}
J~Douglas Carroll, Geert De~Soete, and Sandra Pruzansky.
\newblock Fitting of the latent class model via iteratively reweighted least
  squares candecomp with nonnegativity constraints.
\newblock In \emph{Multiway data analysis}, pages 463--472. North-Holland
  Publishing Co., 1989.

\bibitem[Chandola et~al.(2009)Chandola, Banerjee, and
  Kumar]{chandola2009anomaly}
Varun Chandola, Arindam Banerjee, and Vipin Kumar.
\newblock Anomaly detection: A survey.
\newblock \emph{ACM computing surveys (CSUR)}, 41\penalty0 (3):\penalty0 15,
  2009.

\bibitem[De~Lathauwer et~al.(2000{\natexlab{a}})De~Lathauwer, De~Moor, and
  Vandewalle]{de2000best}
Lieven De~Lathauwer, Bart De~Moor, and Joos Vandewalle.
\newblock On the best rank-1 and rank-(r 1, r 2,..., rn) approximation of
  higher-order tensors.
\newblock \emph{SIAM Journal on Matrix Analysis and Applications}, 21\penalty0
  (4):\penalty0 1324--1342, 2000{\natexlab{a}}.

\bibitem[De~Lathauwer et~al.(2000{\natexlab{b}})De~Lathauwer, De~Moor, and
  Vandewalle]{de2000multilinear}
Lieven De~Lathauwer, Bart De~Moor, and Joos Vandewalle.
\newblock A multilinear singular value decomposition.
\newblock \emph{SIAM journal on Matrix Analysis and Applications}, 21\penalty0
  (4):\penalty0 1253--1278, 2000{\natexlab{b}}.

\bibitem[Fanaee-T and Gama(2015{\natexlab{a}})]{fanaee2015eigenevent}
Hadi Fanaee-T and Jo{\~a}o Gama.
\newblock Eigenevent: An algorithm for event detection from complex data
  streams in syndromic surveillance.
\newblock \emph{Intelligent Data Analysis}, 19\penalty0 (3):\penalty0 597--616,
  2015{\natexlab{a}}.

\bibitem[Fanaee-T and Gama(2015{\natexlab{b}})]{fanaee2015multi}
Hadi Fanaee-T and Jo{\~a}o Gama.
\newblock Multi-aspect-streaming tensor analysis.
\newblock \emph{Knowledge-Based Systems}, 89:\penalty0 332--345,
  2015{\natexlab{b}}.

\bibitem[Fanaee-T and Gama(2016{\natexlab{a}})]{fanaee2016event}
Hadi Fanaee-T and Jo{\~a}o Gama.
\newblock Event detection from traffic tensors: A hybrid model.
\newblock \emph{Neurocomputing}, 203:\penalty0 22--33, 2016{\natexlab{a}}.

\bibitem[Fanaee-T and Gama(2016{\natexlab{b}})]{fanaee2016tensor}
Hadi Fanaee-T and Jo{\~a}o Gama.
\newblock Tensor-based anomaly detection: An interdisciplinary survey.
\newblock \emph{Knowledge-Based Systems}, 98:\penalty0 130--147,
  2016{\natexlab{b}}.

\bibitem[Fanaee-T et~al.(2014)Fanaee-T, Oliveira, Gama, Malinowski, and
  Morla]{fanaee2014event}
Hadi Fanaee-T, M{\'a}rcia~DB Oliveira, Jo{\~a}o Gama, Simon Malinowski, and
  Ricardo Morla.
\newblock Event and anomaly detection using tucker3 decomposition.
\newblock \emph{arXiv preprint arXiv:1406.3266}, 2014.

\bibitem[Hu et~al.(2015)Hu, Rai, Chen, Harding, and Carin]{hu2015scalable}
Changwei Hu, Piyush Rai, Changyou Chen, Matthew Harding, and Lawrence Carin.
\newblock Scalable bayesian non-negative tensor factorization for massive count
  data.
\newblock In \emph{Joint European Conference on Machine Learning and Knowledge
  Discovery in Databases}, pages 53--70. Springer, 2015.

\bibitem[Kim and Choi(2007)]{kim2007nonnegative}
Yong-Deok Kim and Seungjin Choi.
\newblock Nonnegative tucker decomposition.
\newblock In \emph{2007 IEEE Conference on Computer Vision and Pattern
  Recognition}, pages 1--8. IEEE, 2007.

\bibitem[Kolda and Bader(2009)]{kolda2009tensor}
Tamara~G Kolda and Brett~W Bader.
\newblock Tensor decompositions and applications.
\newblock \emph{SIAM review}, 51\penalty0 (3):\penalty0 455--500, 2009.

\bibitem[Kolda and Sun(2008)]{kolda2008scalable}
Tamara~G Kolda and Jimeng Sun.
\newblock Scalable tensor decompositions for multi-aspect data mining.
\newblock In \emph{2008 Eighth IEEE international conference on data mining},
  pages 363--372. IEEE, 2008.

\bibitem[Lemm et~al.(2011)Lemm, Blankertz, Dickhaus, and
  M{\"u}ller]{lemm2011introduction}
Steven Lemm, Benjamin Blankertz, Thorsten Dickhaus, and Klaus-Robert
  M{\"u}ller.
\newblock Introduction to machine learning for brain imaging.
\newblock \emph{Neuroimage}, 56\penalty0 (2):\penalty0 387--399, 2011.

\bibitem[Miettinen(2011)]{miettinen2011boolean}
Pauli Miettinen.
\newblock Boolean tensor factorizations.
\newblock In \emph{2011 IEEE 11th International Conference on Data Mining},
  pages 447--456. IEEE, 2011.

\bibitem[M{\o}rup(2011)]{morup2011applications}
Morten M{\o}rup.
\newblock Applications of tensor (multiway array) factorizations and
  decompositions in data mining.
\newblock \emph{Wiley Interdisciplinary Reviews: Data Mining and Knowledge
  Discovery}, 1\penalty0 (1):\penalty0 24--40, 2011.

\bibitem[M{\o}rup et~al.(2008)M{\o}rup, Hansen, Arnfred, Lim, and
  Madsen]{morup2008shift}
Morten M{\o}rup, Lars~Kai Hansen, Sidse~Marie Arnfred, Lek-Heng Lim, and
  Kristoffer~Hougaard Madsen.
\newblock Shift-invariant multilinear decomposition of neuroimaging data.
\newblock \emph{NeuroImage}, 42\penalty0 (4):\penalty0 1439--1450, 2008.

\bibitem[Nion and Sidiropoulos(2009)]{nion2009adaptive}
Dimitri Nion and Nicholas~D Sidiropoulos.
\newblock Adaptive algorithms to track the parafac decomposition of a
  third-order tensor.
\newblock \emph{IEEE Transactions on Signal Processing}, 57\penalty0
  (6):\penalty0 2299--2310, 2009.

\bibitem[Papalexakis et~al.(2016)Papalexakis, Faloutsos, and
  Sidiropoulos]{papalexakis2016tensors}
Evangelos~E Papalexakis, Christos Faloutsos, and Nicholas~D Sidiropoulos.
\newblock Tensors for data mining and data fusion: Models, applications, and
  scalable algorithms.
\newblock \emph{ACM Transactions on Intelligent Systems and Technology (TIST)},
  8\penalty0 (2):\penalty0 16, 2016.

\bibitem[Sakurai(2016)]{Trimine}
Yasushi Sakurai.
\newblock Trimine.
\newblock Available online, November 2016.
\newblock URL \url{http://www.cs.kumamoto-u.ac.jp/~yasushi/src/trimine.zip}.

\bibitem[Shilon(2016)]{RandOrthMat}
Ofek Shilon.
\newblock Randorthmat.m.
\newblock Available online, November 2016.
\newblock URL
  \url{https://www.mathworks.com/matlabcentral/fileexchange/11783-randorthmat}.

\bibitem[Sun et~al.(2008)Sun, Tao, Papadimitriou, Yu, and
  Faloutsos]{sun2008incremental}
Jimeng Sun, Dacheng Tao, Spiros Papadimitriou, Philip~S Yu, and Christos
  Faloutsos.
\newblock Incremental tensor analysis: Theory and applications.
\newblock \emph{ACM Transactions on Knowledge Discovery from Data (TKDD)},
  2\penalty0 (3):\penalty0 11, 2008.

\bibitem[Tan et~al.(2013)Tan, Wu, Feng, Wang, and Ran]{tan2013new}
Huachun Tan, Yuankai Wu, Guangdong Feng, Wuhong Wang, and Bin Ran.
\newblock A new traffic prediction method based on dynamic tensor completion.
\newblock \emph{Procedia-Social and Behavioral Sciences}, 96:\penalty0
  2431--2442, 2013.

\bibitem[Tucker(1966)]{tucker1966some}
Ledyard~R Tucker.
\newblock Some mathematical notes on three-mode factor analysis.
\newblock \emph{Psychometrika}, 31\penalty0 (3):\penalty0 279--311, 1966.

\bibitem[Tune and Roughan(2013)]{tune2013internet}
Paul Tune and Matthew Roughan.
\newblock Internet traffic matrices: A primer.
\newblock \emph{Recent Advances in Networking}, 1, 2013.

\bibitem[Viswanathan(2015)]{Viswanathan}
Mathuranathan Viswanathan.
\newblock How to generate awgn noise in matlab/octave(without using in-built
  awgn function).
\newblock Available online, June 2015.
\newblock URL
  \url{http://www.gaussianwaves.com/gaussianwaves/wp-content/uploads/2015/06/How_to_generate_AWGN_noise.pdf}.

\bibitem[Zhao et~al.(2016)Zhao, Zhou, Zhang, Cichocki, and
  Amari]{zhao2016bayesian}
Qibin Zhao, Guoxu Zhou, Liqing Zhang, Andrzej Cichocki, and Shun-Ichi Amari.
\newblock Bayesian robust tensor factorization for incomplete multiway data.
\newblock \emph{IEEE transactions on neural networks and learning systems},
  27\penalty0 (4):\penalty0 736--748, 2016.

\bibitem[Zhou et~al.(2016)Zhou, Vinh, Bailey, Jia, and
  Davidson]{zhouaccelerating}
Shuo Zhou, Nguyen~Xuan Vinh, James Bailey, Yunzhe Jia, and Ian Davidson.
\newblock Accelerating online cp decompositions for higher order tensors.
\newblock In \emph{Proceedings of the 22Nd ACM SIGKDD International Conference
  on Knowledge Discovery and Data Mining}, KDD '16, pages 1375--1384, New York,
  NY, USA, 2016. ACM.
\newblock ISBN 978-1-4503-4232-2.
\newblock \doi{10.1145/2939672.2939763}.
\newblock URL \url{http://doi.acm.org/10.1145/2939672.2939763}.

\end{thebibliography}

\end{document}